\documentstyle{aipproc}

\newcommand{\lapprox}{\stackrel{<}{\scriptstyle \sim}}

\input epsf
\begin{document}
\title{(Hybrid) Baryons in the Flux--Tube Model}

\author{Philip R. Page\thanks{prp@jlab.org. Work done in collaboration
with Simon Capstick. Contribution to 
the Seventh International Conference on Hadron Spectroscopy
(HADRON '97), Brookhaven, August 1997. I acknowledge a Lindemann Fellowship from the English Speaking Union.}$^*$}
\address{$^*$ Theory Group, Thomas Jefferson National Accelerator Facility,\\
12000 Jefferson Avenue, Newport News, VA 23606, USA }

%\lefthead{LEFT head}
%\righthead{RIGHT head}
\maketitle

\begin{abstract}
We construct baryons and hybrid baryons in the
non--relativistic flux--tube model of Isgur and Paton. The motion of
the flux--tube with the three quark positions fixed, except for centre
of mass corrections, is discussed. It is
shown that the problem can  to an excellent approximation be reduced to the independent motion of a
junction and strings. 
\end{abstract}

Hybrids are bound states where there is an explicit excitation in the
gluon field of QCD. Particularly, ``hybrids baryons'' may be viewed as
quark--quark--quark--glue composites. 

The experimental interest in hybrid baryons centers around the excited
baryon resonance ($< 2.2$ GeV) program at TJNAF, mostly in Hall
B. Hybrid baryon production is expected. If hybrid baryons obey
similar decay selection rules to hybrid mesons, they may be
distinguishable based on their strong decays.

Hybrid baryons have only been constructed in the MIT bag model
\cite{bag}. We are motivated to build a model consistent\footnote{The
hybrid meson interquark potential is consistent with that evaluated from
lattice gauge theory \protect\cite{paton85}.} with
predictions from QCD lattice gauge theory, i.e. the Isgur--Paton
non--relativistic flux--tube model \cite{paton85}. This model is
motivated from the strong coupling limit of the hamiltonian lattice
gauge theory formulation of QCD (HLGT).

This talk will deal with fixed quark positions relative to each
other. However, we shall allow the quarks with fixed relative
positions to move in order to work in the centre of mass frame. This
is called the ``quasi--adiabatic'' approximation. 

The model is motivated from the strong coupling limit of HLGT, where there are
``flux--lines'' which play the role of glue.
In the spirit of the adiabatic approximation, where quarks do not
respond to the influence of glue, we neglect 
operators which make quarks move. The remaining operator taking you
away from the strong coupling limit is the plaquette operator. This
induces motion of the ``flux--line'' between the quarks perpendicular
to its rest position. We model the flux--line by ``beads'', all with the
same mass, which is fixed from the energy in the linear flux--line. The
beads are seperated along their rest positions by a finite lattice spacing, and are allowed to
move perpendicular to their rest position. The beads are attracted to
each other by a linear potential, vibrating in various
string modes. 

There is a second essential ingredient from HLGT. Three flux--lines
can come together at a point called the ``junction''. A
plaquette operator cannot move the junction in the lowest order of
perturbation theory so that the junction is taken to have a different
(higher) mass associated with it than the other beads.

The final picture of a (hybrid) baryon is that of three quarks, each
connected via a line of beads to the junction in a Mercedes Benz configuration. 
The three quarks define a plane. The ``equilibrium configuration'' is
the lowest energy configuration: the junction is located such that there are
angles of $120^o$ between each of the ``triads'' that connect each of
the quarks to the junction, and the beads all lie on the triads. The
junction and beads then vibrate with respect to the equilibrium
configuration. There are two important motions which are expected to
have physical significance: (1) the motion of the junction
perpendicular and within the plane relative to the junction rest
position, called the ``junction motion''; (2) the motion of the beads
in the two directions perpendicular to the line connecting the quark
to the junction, called the ``string motion''.
If the angles between two of the quarks suspended at the third quark
is larger than $120^o$ the equilibrium configuration is not the
Mercedes Benz configuration. This issue is not considered further here.

We now make the small oscillation approximation, where the beads and
junction move near to the equilibrium configuration. We make sure that
we work in the centre of mass system, and therefore make the
``quasi--adiabatic'' approximation. The hamiltonian is written in
terms of the junction and string motion coordinates. 

We have demonstated that the hamiltonian can be seperated into a
part $H_J$
which corresponds to the motion of the junction in the potential one
would use if there were no beads in the problem, with an effective
junction mass
related to its own mass and the mass of the beads. Another part $H_S$ is the
independent motion of three strings with respect to a fixed junction, with an effective
bead mass that is related to the bead and junction masses. There is
also an ``interaction term'' $H_{int}$ where the strings corresponding to
different quarks interact with each other and various string modes 
associated with the same quark interact with each other, and where
the junction interacts with the various string modes.

We shall now demonstate that the interaction term gives a minor contribution. 
The free parameters in the model (and the values used for the numerical
simulation) are the string tension ($0.18$
GeV$^2$), the ratio of the junction and bead masses (1) and the quark
masses ($0.33$ GeV). We shall perform a simulation where there is one
bead between each quark and the junction and the quarks
form an equilateral triangle with the lengths of the triads equal to a
typical value of $2.5$ GeV$^{-1}$. 

First we solve the exact problem numerically. The frequencies parallel
and perpendicular to the plane are (in GeV)

\begin{tabbing} \label{bnl2}
XXXXXXXXXXXXXXXX\=XXXXXX\=XXXXXX\=XXXXXX\=XXXXXX\=XXXXXX\kill 
Parallel\>{\bf 0.607}\>{\bf 0.607}\>0.924\>1.08\>1.08\\
Perpendicular\>{\bf 0.828}\>\>0.924\>0.924\>1.37\\
\end{tabbing}
where the bold faced frequencies are clearly lower than the
others. In fact, if we set $H_{int}=0$ then we again obtain the same
number of frequencies, and here quote the results for the lowest lying
frequencies corresponding to the junction motion (in GeV):

\begin{tabbing} \label{bnl2}
XXXXXXXXXXXXXXXX\=XXXXXX\=XXXXXX\=XXXXXX\=XXXXXX\=XXXXXX\kill 
Parallel\>{\bf 0.614}\>{\bf 0.614}\\
Perpendicular\>{\bf 0.869}\\
\end{tabbing}
We hence conclude that the lowest frequencies are the ones
corresponding to the junction motion, and that we can neglect
$H_{int}$ safely for the lowest frequency. In retrospect, the reason why $H_{int}$ can be
neglected is because we chose the physically appropriate coordinates
for the problem: the junction and string motions.

To compare the frequencies of the full hamiltonian and
the junction frequencies of the hamiltonian with $H_{int}=0$, we calculate the
deviation $\epsilon$ of them from one another. $\epsilon_{-}= 1\%$ and
$\epsilon_{+}= 1\%$ for the low and high parallel frequencies respectively
(in this case the frequencies are equal). $\epsilon = 5\%$ perpendicular
to the plane. For the lowest frequency, 
it is hence sufficient to work with a hamiltonian of the
form $H_J+H_S$ from now on, and there are three types of hybrid
baryons: the one corresponding to junction motion perpendicular to the plane
which is always the heaviest; and two corresponding to motion parallel
to the plane. For a generic quark configuration, one of the parallel
frequencies is always below or equal to the other, so that they are
not usually degenerate. The flux--tube model thus contains three low lying
hybrid baryons, corresponding to vibrations along three perpendicular
axes, but each with a different excitation frequency above the baryon.

To access the error in the hamiltonian  with $H_{int}=0$ more fully,
we vary parameters around the central values above, one at a time; with the quark mass up to
the charm quark mass of $1.5$ GeV, the ratio of the junction to the bead
mass up to 10, and with triads with lengths from $0.5 - 5$ GeV$^{-1}$
(the triad
lengths not being equal in general). We found that that $\epsilon_{-} \lapprox
5\%$ and $\epsilon_{+} \lapprox
6\%$ for parallel frequencies and $\epsilon \lapprox 40\%$ for the perpendicular
frequency. The error introduced by neglecting $H_{int}$ is therefore
rather minimal for the lowest frequency. Hence, to a good approximation, the dynamics of the lowest frequency can be
simplified to junction and string motion which are independent of one
another. 

Significant progress has been made towards building a realistic
flux--tube model of (hybrid) baryons.
We have constructed the full multibead hamiltonian in the
quasi--adiabatic approximation and in the small oscillations approximation. We demonstated that the junction bead
decouples from the other beads to a high degree of accuracy.


\begin{references}
\bibitem{bag} Barnes, T. and Close, F.E., {\it Phys. Lett.} {\bf
B123}, 89 (1983); Golowich, E., Haqq, E. and Karl, G., {Phys. Rev.} {\bf D28}, 160
(1983); Carlson, C.E. and Hansson, T.H., {\it Phys. Lett.} {\bf B128}, 95 (1983);
Carlson, C.E., BARYON '95 (1995).
\bibitem{paton85} Isgur, N. and Paton, J., {\it Phys. Rev.} {\bf D31}
(1985) 2910.
%\bibitem{cap} Capstick, S. and Page, P.R., {\it in preparation}.





\end{references}
\end{document}